\documentclass[conference]{IEEEtran}
\IEEEoverridecommandlockouts
\usepackage{cite}
\usepackage{amsmath,amssymb,amsfonts}
\usepackage{algorithmic}
\usepackage{graphicx}
\usepackage{textcomp}
\usepackage{a4wide}
\usepackage{array}
\usepackage{hyperref}
\usepackage{amssymb}
\usepackage{amsmath}
\usepackage{color}
\usepackage[table]{xcolor}
\usepackage{multirow}
\usepackage{enumitem}
\usepackage{float}
\usepackage{graphicx}
\usepackage{hhline}
\usepackage{listings}
\usepackage{lscape}
\usepackage{rotating}
\usepackage{subcaption}
\usepackage{url}
\usepackage{wrapfig}
\usepackage{xargs}  
\usepackage{pgf-pie}
\usepackage{pgfplots}
\usepackage{siunitx}
\usepackage{stfloats}

\def\BibTeX{{\rm B\kern-.05em{\sc i\kern-.025em b}\kern-.08em
    T\kern-.1667em\lower.7ex\hbox{E}\kern-.125emX}}
\begin{document}

\title{Evaluation of Multi-Scale Multiple Instance Learning to Improve Thyroid Cancer Classification\\
\thanks{This work was partially funded by the County of Salzburg under grant number FHS2019-10-KIAMed and by the Austrian Agency for International Cooperation in Education and Research (OeAD-GmbH) HR02/2018.}
}

\author{\IEEEauthorblockN{Maximilian E. Tschuchnig\IEEEauthorrefmark{1}\IEEEauthorrefmark{4}, Philipp Grubmüller\IEEEauthorrefmark{1}, Lea M. Stangassinger\IEEEauthorrefmark{1}, Christina Kreutzer\IEEEauthorrefmark{2},\\ Sébastien Couillard-Després\IEEEauthorrefmark{2}, Gertie J. Oostingh\IEEEauthorrefmark{1}, Anton Hittmair\IEEEauthorrefmark{3}, Michael Gadermayr\IEEEauthorrefmark{1}
    \IEEEauthorblockA{\IEEEauthorrefmark{1}Salzburg University of Applied Sciences, maximilian.tschuchnig@fh-salzburg.ac.at}
    \IEEEauthorblockA{\IEEEauthorrefmark{2}Spinal Cord Injury and Tissue Regeneration Center Salzburg,\\Research Institute of Experimental Neuroregeneration}
    \IEEEauthorblockA{\IEEEauthorrefmark{3}Kardinal Schwarzenberg Klinikum, Department of Pathology and Microbiology} 
    \IEEEauthorblockA{\IEEEauthorrefmark{4}University of Salzburg, Department of Artificial Intelligence and Human Interfaces} 
    }
}

\maketitle

\begin{abstract}
Thyroid cancer is currently the fifth most common malignancy diagnosed in women. Since differentiation of cancer sub-types is important for treatment and current, manual methods are time consuming and subjective, automatic computer-aided differentiation of cancer types is crucial. Manual differentiation of thyroid cancer is based on tissue sections, analysed by pathologists using histological features. Due to the enormous size of gigapixel whole slide images, holistic classification using deep learning methods is not feasible. Patch based multiple instance learning approaches, combined with aggregations such as bag-of-words, is a common approach. This work's contribution is to extend a patch based state-of-the-art method by generating and combining feature vectors of three different patch resolutions and analysing three distinct ways of combining them. The results showed improvements in one of the three multi-scale approaches, while the others led to decreased scores. This provides motivation for analysis and discussion of the individual approaches.
\end{abstract}


\vspace{-1em}
\section{Introduction}

Thyroid cancer is currently the fifth most common malignancy diagnosed in women ($19$th most common in men)\cite{siegel2020} with the most common sub-types being papillary cancer (PC) and follicular nodule (FN)\cite{siegel2020}. To diagnose thyroid cancer, fine-needle aspiration (FNA) is performed. If diagnosis was not possible by FNA, a surgical biopsy is advised \cite{tuttle2010thyroid}. The extracted tissue is then prepared for analysis by embedding it in paraffin, making the tissue compatible with a variety of staining methods and allowing for thin sectioning. Frozen sections are a further method of tissue preparation, typically generated quickly during interventions, however, the quality is lower compared to paraffin sections \cite{gadermayr2021}. To enable digital processing, these prepared slices can be digitized using Whole Slide Image (WSI) scanners, enabling digital processing and the application of e.g. Machine Learning and particularly Deep Learning (DL) models.

Intraobserver variation in the diagnosis of thyroid cancer is high, especially for PC \cite{hirokawa2002}. This effect is amplified by an incomplete and therefore partially subjective definition of thyroid cancer\cite{hirokawa2002}. Differentiation is important to establish fitting treatment \cite{shah1992}, especially to establish the extent of surgery and of radiodine and thyroid suppression therapy. Due to the increasing availability of whole slide scanners, computer-aided diagnosis (CAD) can be used to automate cancer differentiation in an objective way. Using CAD to establish automated cancer differentiation is further motivation by DL based CAD reaching comparable scores to pathologists \cite{duran2020, li2018} or in some cases even outperforming pathologists (in narrow fields of applications and under time pressure) \cite{bejnordi2017, miyoshi2020}. A partially or fully automated decision support system would aid pathologists in diagnostics, with the potential of reducing diagnosis time and increasing cancer sub-type differentiation accuracy, leading to an increase in treatment efficiency and patient safety. 

Digital whole slide images (WSI) enable a wide range of applications, among others they enable CAD. However, due to their size, the gigapixel WSIs cannot be directly processed by most DL methods, such as convolutional neural networks (CNN). Due to the assumption that cancer sub-type differentiation is based on image differences on a cellular level \cite{hou2016}, downsampling to a (for CNNs) usable format is ineffective. 

One solution to this issue is to apply multiple instance learning (MIL) \cite{hou2016}. In the MIL paradigm, unlabelled instances belong to bags of instances with the goal of labelling these bags with a positive label as soon as one instance is labelled as positive. % The Standard Multi-Instance (SMI) assumption \cite{dietterich1997} states that in binary classification, a bag is positive if at least one instance in the bag is positive. Using neural network this decision is typically replaced by max-pooling. % Not really needed
In digital histology, count-based MIL (CMI) has been shown to be successful \cite{hou2016, marini2021, gadermayr2021}. CMI is a more general approach to MIL, using an undefined aggregation operation to combine the different instances. This allows e.g. for subsequent classification using support vector machines (SVM).

\textit{Contribution: } In this paper multiple multi-scale MIL approaches differentiating FN and PC were evaluated and compared to a single resolution MIL baseline originally proposed in \cite{gadermayr2021}. Classification accuracy was improved in case of one setting whereas scores decreased with the other investigated settings. It was further discussed how multi-scale approaches impact classification accuracy and future research directions were proposed.

% Only image level ground truth is given % Is implied enough

\section{Related Work}

Hou et al. \cite{hou2016} introduced a patch-based CNN approach for WSI cancer classification using CMI. Since the discriminative information is encoded in high resolution images, the chosen approach was to predict the label based on aggregated patch predictions \cite{hou2016}. Logistic regression as well as SVMs were used as a MIL decision fusion model to aggregate the patch level predictions and predict the WSI label. % Only the image-level ground truth is given. Important fusion problem: voting and max-pooling are not robust since cancer can be very local, with distinct regions for possible cancer sub-types. This issue was approached by eliminating non discriminative patches using an iterative EM-based method. This allows simple max-pooling or voting or more sophisticated predictions using multiclass logistic regression or SVMs.

% The paper by Gadermayr et al. \cite{gadermayr2021} uses a patch based classification method, similar to CMI. In their work they introduce a new thyroid gland carcinoma dataset and, among others, differentiate between paraffin PC and FC carcinomas. Their method ... % We do not need our own paper here right? we talk about it in detail on multiple occasions

Li et al. \cite{li2021} also formalize WSI classification as a MIL problem and expanded on this by introducing a novel deep MIL model, dual-stream MIL. This dual-stream model was extended to include differently scaled feature patches with one concatenated multi-scaled patch feature consisting of $4$ (patch) feature vectors with $5 \times$ magnification, $16$ with $10 \times$ magnification and $64$ with $20 \times$ magnification. These individual feature vectors were concatenated in the same way as a corresponding image pyramid. Since the different magnification levels consisted of a different amount of patches, the lower resolution patch feature vectors were repeated to fit the size of the $20 \times$ magnification vector. This multi-scale approach was motivated by the idea of leveraging both millimeter scale features (vessels and glands) and cellular scales (tissue microenvironment). However, due to their proposed aggregation and extraction of an increased number of patches on higher resolutions, the positive effect of their contribution cannot conclusively be attributed to the use of multi-scale patches. This is also reported in the paper, since it is remarked that using $2$ magnification levels reaches better scores than all $3$. In order to better evaluate the effect of multi-scaled patches, using only a single, corresponding patch per magnification level is preferred.

A multi-scale approach that uses only one patch per scale was proposed by Marini et al. \cite{marini2021}. They also provide an adaptation of MIL denoted as Multi-Scale Task MIL. Using their proposed methods they were able to outperfom baseline WSI classifcation MIL algorithms, however, similarly to \cite{li2021}, the multi-scale approach is only a minor improvement in some cases, compared to their single-scale approach. Further investigation of the applicability of multi-scaled approaches are therefore necessary.

\section{Dataset}

The dataset used was the same as in previous work \cite{gadermayr2021} and consisted of $80$ WSIs of different image modalities (frozen and paraffin) from which only paraffin-treated WSIs were used, reducing the dataset to $40$ WSIs. All images were acquired during clinical routine at the Kardinal Schwarzenberg Hospital.
The WSIs were generated in the following way. Paraffin sections were fixed in $4\%$ phosphate-buffered formalin for $24$ hours. The formalin fixed embedded tissue was cut ($2 \mu m$) and stained with hematoxylin and eosin. The images were then digitized with an Olympus VS120-LD100 slide load scanner with $2$ times magnification overviews for scan area definition and $20$ times magnification ($344.57 nm/pixel$) for the WSI data. This data was stored in the lossless Olympus vsi format and labelled by an expert pathologist with over $20$ years experience. A total of $21$ slides were labeled as PC while $19$ were labeled as FN. For the purpose of anonymization, the patients were adapted with an anonymous patient-id. % The chosen method aimed to mimic the procedure in which pathologists analyse histological cuts for thyroid carcinoma by first focussing on a general area-of-interest and then taking a more detailed look. This, importantly also investigates if differentiating features are only present in the highest ($20x$) resolutions. This increase in scale was performed twice, leading to three different patch resolutions, covering increasing areas of the WSI (Fig.~\ref{Fig:PatchExtraction}). % Is this even needed?

\section{Method}

In this work, we aimed to improve the classification of different nodular lesions of the thyroid, FN and PC as proposed in \cite{gadermayr2021}. % This differentiation is crucial, since it leads to target specific therapies. % Redundant
To accomplish this, we built on \cite{li2021, gadermayr2021, marini2021} and applied a multi-scale MIL approach. To focus only on the effect of different multi-scale approaches, we rely on a basic overall architecture \cite{gadermayr2021} and keep the majority of parameters unchanged, investigating only a small set of parameters and the effect of using multiple scales. As a baseline we employ the method introduced in \cite{gadermayr2021}. In the baseline, patches were extracted from WSIs, which were then applied to ResNet18, generating feature vectors per patch. These feature vectors were then clustered and a histogram, aggregating all patch clusters per WSI was generated (bag-of-words). These bags-of-words were used to train and evaluate different SVMs, with self optimizing SVMs as well as SVMs using linear and RBF kernels. The optimized SVM used inner cross validation with the tuning parameters $\gamma = {1e^{-3}, 1e^{-4}}$ and cost factor $c = {0.5, 1, 2, 4, 8, 16, 32}$ with linear and RBF kernels. The multi-scale approach builds on the baseline by additionally extracting multi-scaled patches, which again were applied to ResNet18 for feature generation. SVMs were used to classify the WSIs after three different methods of feature aggregation.

\begin{figure}[htp]
	\begin{center}
	\fbox{\includegraphics[width=0.95\linewidth]{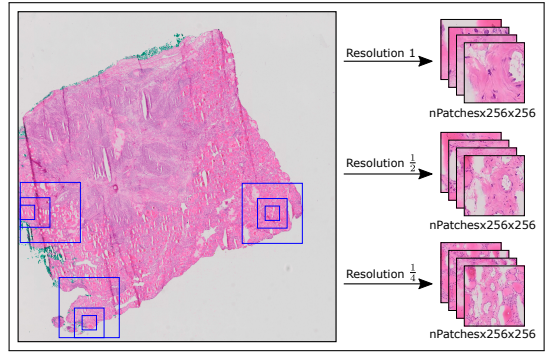}}
	\end{center}
	\caption{Multi-scale patch extraction process. Starting points of the highest resolution patch were uniformly sampled from the WSI dimensions. This is followed by checks on if the patches show tissue and if the lower resolution patches (which were spatially larger) fit inside the WSI. If they did not, the starting positions for the current patch was moved by $\frac{1}{4}$ of its size towards the image center. This movement was done for both axis.
	\vspace{-1em}}
	\label{Fig:PatchExtraction}
\end{figure}

\textit{Patch Extraction:} % Digital thyroid carcinoma classification is based on gigapixel WSIs which are challenging to work with using e.g. CNNs. Therefore, patch based MIL methods are applied, extracting multiple, randomly selected patches from the WSIs with the aim of capturing all the important information, aggregating this information and finally using a count based fusion method to obtain a global image label. % Redundant
Fig. \ref{Fig:PatchExtraction} shows the process of extracting random, multi-scaled patches. First, a random position was chosen as the patch starting point within the image boundaries. The green image color channel was used to check if the patch was showing tissue by only accepting mean values lower than $190$ (experimentally chosen). The green color channel was chosen due to the effectiveness of separating the background from the tissue using thresholds. To establish the multi-scale feature extraction, this process is adapted. Similar to before, the smallest scaled patch position is randomly chosen and checked for validity with the introduced threshold. Then patches with halved and quartered resolutions were generated, doubling in size with each halved resolution, with the original scaled patch positioned in their middle. If this increase in size leads to the lower scaled patches positioned partially outside the WSI (e.g. the two multi-sclaed patches of Fig.~\ref{Fig:PatchExtraction} in the bottom left), the lower scaled patches were moved by $\frac{1}{4}$ their size towards the image center. Finally, the patches were downscaled to the same size ($256 \times 256$) and stored. % Additionally, the patch positions (top-left pixel) of the highest scales were stored for future explainability analysis. % future paper

\begin{figure*}[!ht]
	\begin{center}
	\fbox{\includegraphics[width=1.\linewidth]{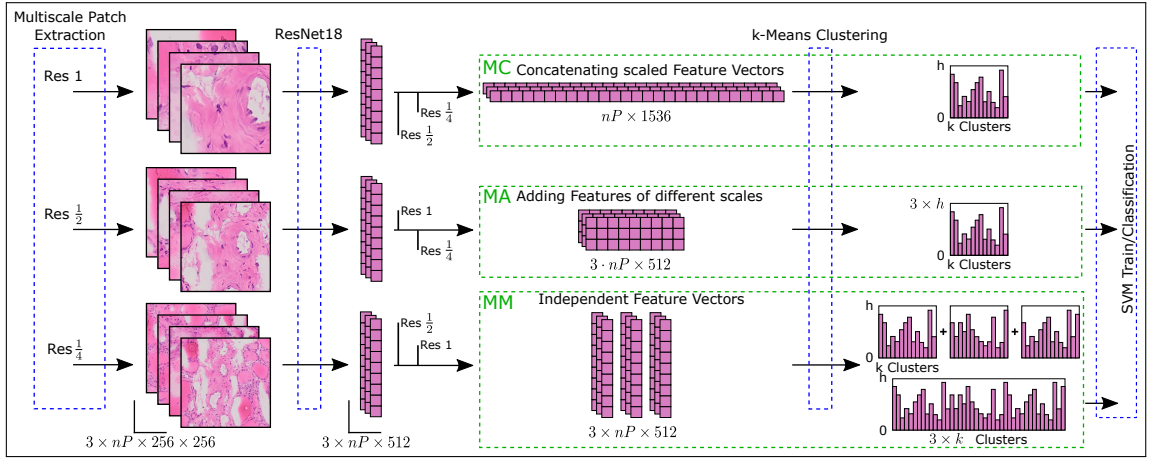}}
	\end{center}
	\caption{This figure shows the different multi-scale cancer differentiation pipelines% , namely Multi-scale Concat (MC), Multi-scale Adapted (MA) and Multi-scale Multiple-k-means (MM). 
	. After extracting multi-scale patches, features are extracted from these patches by ResNet18 which are uniquely aggregated in MC and MA. MM handles the features of the different scales independently. In MC and MA, these features are then clustered per patch and these clusters are finally aggregated into one bag-of-words per WSI. MM clusters the features of each scale independently and aggregates them into three different histograms. MM then concatenates these bag-of-words which are finally used to train/evaluate SVMs.
	\vspace{-1em}}
	\label{Fig:Pipeline}
\end{figure*}

\textit{Classification Pipeline:}
Classification followed the patch-based CMI pipeline introduced in \cite{gadermayr2021} improved by including the additional multi-scale patches. WSI feature augmentation was also applied in the same way as in \cite{gadermayr2021} introduced as "Aug1", which reduced the amount of sampled patches to $75\%$. This reduction is repeated $8$ times, leading to an artificial increase in the dataset. The classification pipeline is shown in Fig. \ref{Fig:Pipeline}. After extracting multiple patches (amount of extracted patches $ = nP$) of three different scales, features from these patches were extracted using ResNet18 resulting in a $(1,512)$ dimensional feature vector per patch and scale. An image-net pre-trained ResNet18 was applied as a feature extractor because it performed well in previous work \cite{dimitriou2019, gadermayr2021, li2021, marini2021}. Three different methods of combining these features per scale were investigated, Multi-scale Concat (MC), Multi-scale Adapted (MA) and Multi-scale Multiple-k-means (MM). Each of these methods used features from all scales (${1,\frac{1}{2},\frac{1}{4}}$). These combined features were then clustered, resulting in one out of $k$ clusters per patch. As a clustering algorithm for all of the three methods, k-means, with the Euclidian distance metric and varied cluster center amounts ($k \in \{32,64,128,256\}$) was chosen. These clusters were aggregated into bags-of-words per WSI. The resulting bags-of-words of all three methods, as well as the baseline method, were used to train and evaluate classification SVMs. For the SVM, linear and RBF kernels were evaluated. Further an internally optimized SVM was evaluated.

For the first method, MC, we considered the concatenation of features of the different scales, generating a single, $1 \times 3 \cdot 512$ sized feature vector per patch (Fig.~\ref{Fig:Pipeline} MC), similar to the approaches of Marini et al. and li et al. \cite{marini2021, li2021}. Each of these feature vectors were clustered into $k$ clusters per patch and aggregated into a $k$ wide histogram per WSI. The reasoning behind concatenating the features on a patch level was to increase the information per patch (for the clustering algorithm). 
The second method (MA) handles the differently scaled patches as patches of the same scale, increasing the feature vector amount per WSI to $3 \cdot nP \times 512$ without changing the size of the feature vectors (Fig.~\ref{Fig:Pipeline} MA). %This effectively assumes that the same features are differently pronounced at different scales and that this differentiation is useful for further classification. 
These feature vectors were clustered by a single k-means algorithm resulting in $3 \cdot nP$ clustering points which were aggregated into a k-wide histogram. The last method (MM) applied individual k-means instances for each scale and aggregated the resulting histograms (Fig.~\ref{Fig:Pipeline} MM), leading to a histogram of width $3 \cdot k$. This method is based on the assumption, similarly to MC, that the features of the different scales should be handled independently while keeping the feature dimensions for clustering at the same size as the baseline ($nP \times 512$). 

This pipeline was separated into a training and an evaluation part. The dataset was randomly split into a train and evaluation dataset with the $40$ WSIs separated into $80\%$ for training and $20\%$ for testing. During training, the different feature vectors were generated and the k-means clustering algorithm(s) were trained (depending on the method, multiple k-means may have had to be trained). The trained models were then applied to the evaluation data and the resulting mean accuracies and standard deviations were reported. Accuracy was chosen as a fitting metric since it is an interpretable metric and the dataset was balanced. This whole process was repeated $512$ times to ensure stable results. In order to compare the different methods to the baseline, multi-scaled patches were extracted and the baseline as well as the three different methods were evaluated using the introduced pipeline and different amounts of k-means clusters ($k \in \{32, 64, 128, 256\}$). These clusters were chosen since $32$ and $64$ showed the best results in the baseline paper and due to the assumption, that more clusters are needed to handle an increase in complexity of the datasets, introduced by the different scales.

\section{Results}

\begin{figure*}[!ht]
	\begin{center}
	\fbox{\includegraphics[width=1\linewidth]{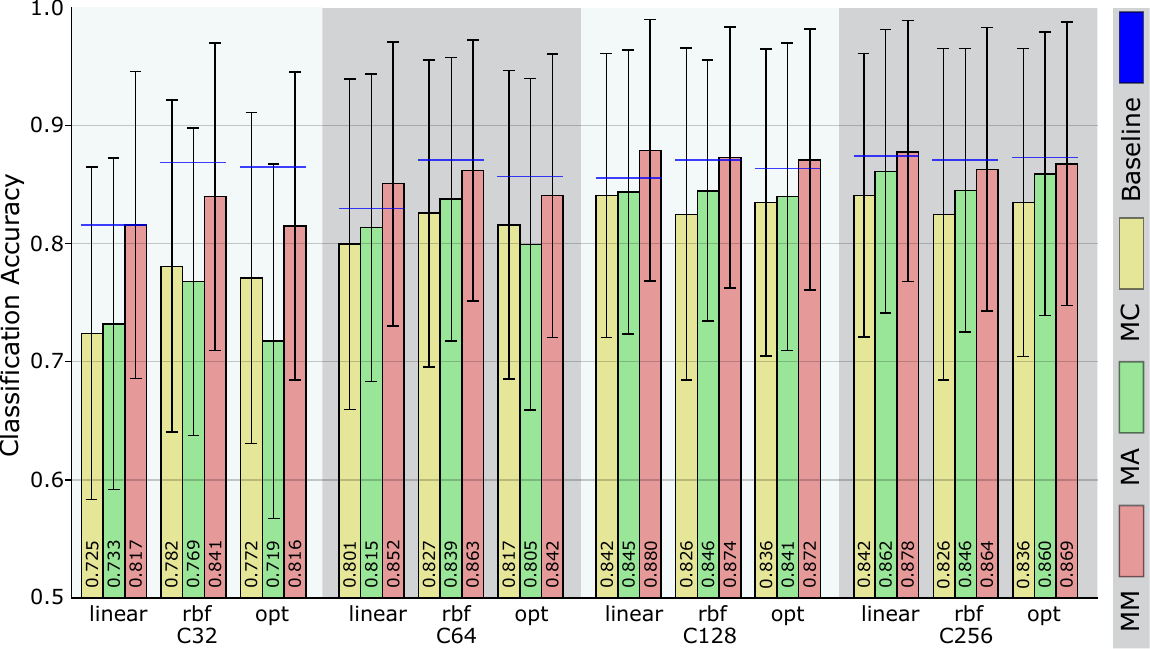}}
	\end{center}
	\caption{This figure shows the results of the experiments with the x-Axis denoting the different numbers of clusters and methods and the y-Axis denoting the classification accuracy starting from $0.5$. The plots show the different methods with the same setup ($svm: {linear, RBF, optimized}$, $k = {32, 64, 128, 256}$). The mean baseline accuracy is displayed as horizontal blue lines. The bars correspond to the models reached mean accuracy with the bar color denoting the experiment (MC: yello, MA: green, MM: red). The black lines on top of the bars correspond to the standard deviation of the current bar.
	\vspace{-1em}}
	\label{Fig:Results}
\end{figure*}

The results of the experiments showed an improvement compared to the baseline using the MM method with $128$ and $256$ k-means clusters from a mean accuracy (acc) of $0.872$ to $0.88$ and a standard deviation (std) of $0.12$ to $0.11$. The highest reported MC results were an $\overline{acc} = 0.842$ with a $std = 0.14$ ($k = \{128, 256\}$). MA had a highest reported $\overline{acc} = 0,862$ with a $std = 0,12$ ($k = 256$) and MM a highest reported $\overline{acc} = 0.88$ and a $std = 0.11$ ($k = 128$). The baseline was reaching a maximum $\overline{acc} = 0.875$ with an $std = 0.12$ ($k = 256$). Both MC and MA did not reach scores higher than the corresponding baselines. Only MM was able to reach higher scores with linear kernel SVMs, scoring $\overline{acc} = 0.88$ and a $std = 0.11$ ($k = \{128, 256\}$). Fig.~\ref{Fig:Results} shows the results with the x-axis denoting the different numbers of clusters and methods and the y-axis denoting the classification accuracy starting from $0.5$ for improved resolution. The plots show the different methods with the chosen SVM setups (linear, RBF, optimized) and the $4$ different chosen number of cluster centers ($k = \{32, 64, 128, 256\}$). The mean baseline accuracy is displayed as horizontal blue lines. The bars correspond to the models reached mean accuracy with the bar color denoting the experiment (MC: yellow, MA: green, MM: red). The black lines on top of the bars correspond to the standard deviation of the current bar.

% \textbf{UPDATE THIS STILL CHANGES, MAYBE BETTER AS GRAPH} The baseline showed no considerable observable trend by increasing the cluster centers $change = {+0.014, +0.026, +0.018}$. MC and MA show a strong trend of accuracy improvement ($\delta MC = {+0.076, +0.041, +0.004}$, $\delta MA = {+0.082, + 0.03, + 0.17}$) by increasing cluster centers, this however flattens, reaching between $128 - 256$ cluster centers. MM shows less of a trend in accuracy improvement through cluster center increase $change = {+0.035, 0.028, 0}$. 

Since the trend was increasing for the baseline, $512$ cluster centers were investigated (baseline and MM), leading to maximum baseline scores of $\overline{acc} = 0.874$ and $std = 0.12$ and an MM $\overline{acc} = 0.876$ and a $std = 0.11$. % !This leads us to the assumption that the optimum of MM cluster centers lies between $128$ and $256$.% No Assuming in results! % ... Removed, not needed ... shows the trends of the classification accuracy, depending on the amount of cluster centers. It shows stable results for the baseline among all $k$ except when a linear SVM is chosen. Overall the linear SVM is shown to increase the most through an increase of cluster centers. MC, MA and MM all showed that with an increase in cluster centers, the RBF kernel SVM models accuracy levels of or decreases. The optimized svm based models accuracy increased using more cluster centers, however it lacked behind the increase of linear svm based models.

\section{Discussion}

The results showed a decrease in accuracy using MC and MA, while MM improved classification accuracy, combined with a lowered standard deviation. However, these improvements of MM were comparably minor, which was also the case in \cite{li2021, marini2021}. One reason for this may be that the most important information for FN and PC differentiation are most prominent in the highest scale as assumed in \cite{hou2016} and that the larger but lower resolution features were less important for classification. If this was true, MC would generate feature vectors with important features on the highest scales but unimportant, and therefore noisy features in the lower scales, leading to a vector, tripled in size for k-means to process with $\frac{2}{3}$ noise. This decreases the possibility for k-means to cluster efficiently, especially using equal or fewer cluster centers. This noise could be one way to explain why the baseline scores could not be reached. Since in MA, the same k-means algorithm is applied to all three different resolutions of patches, aggregating the histograms will effectively lead to adding noise on top of each histogram. This effect could vanish with an increase in the number of cluster centers by clustering noisy data into specific clusters. These noise clusters could then be ignored by the SVM, leading to baseline scores, however this behaviour could not be observed. Both accuracies were shown to increase with an increase in $k$, however MC ad MA $\overline{acc}$ never reached baseline $\overline{acc}$. MM used the additional features, by generating different k-means models for each scale. By concatenating the assumed noisy histograms before training/evaluating the SVM, noise remains. Since the results showed higher classification accuracy than the baseline using MM and $k > 64$ we assume that useful information could be extracted, mitigating the noise. This paper therefore suggests that the most important features for FN and PC differentiation are mostly present at the highest resolutions and MM was able to extract useful information from the lower scales. It is important to remember that, while the lower resolutions have less cellular level information, they have a bigger area of interest.

Another question is how more discriminative information on higher scales would influence the performance of the introduced aggregation methods. The first major differentiation is that MC is higher dimensional in k-means clustering while MM moves the higher dimensionality to the SVM, which typically can handle dimensionality better than k-means. This leads to MM more optimally using independent, multi-scale features. However, if features are not independent and at least partially overlap (e.g. as the same feature with a different scale), MM has increased difficulties handling these relationships. MC and MA can use these relationships already in the clustering step, possibly improving results. MA is very interesting under these aspects since, assuming k is increased sufficiently, a separation of clusters similar to MM should be reached while still keeping relationships between scales. Since MC struggles with high amounts of features in the clustering step it would be of interest to also investigate a feature extractor with less output features than the $512$ of ResNet18 or with an initial step of dimensionality reduction.

% These results showed that a multi-scale approach could lead to classification improvements, however there are relevant biases that have to be taken into account. One adaptation that would be to, instead of using multi-scaled patches, using multiple neighbour patches of different resolutions and combining them in the same way as MA as an additional data augmentation step. From the results of this evaluation we conclude that, replacing the multi-scale approach with multiple patches of the highest resolution shows high potential.

It is also important to note, that we observed bias depending on the patch extraction which was due to the small sample size. A further important note is that the reported results depend on the tissue to be analyzed as in e.g. \cite{marini2021}, the most promising magnification level to detect colon cancer was not $20 \times$ as it was for thyroid cancer but $5 \times$.

\section{Conclusion}

This paper investigated three different multi-scale MIL approaches, classifying thyroid cancer into sub-classes (FN and PC) and comparing them to a baseline. Three intuitive MIL approaches were evaluated with variable settings regarding the composition of the bag-of-words and the classification method. We showed that the classification into sub-types was successful using MM (reaching mean accuracies of $0.88$) whereas MA and MC decreased classification performance. 

In future experiments an investigation of multi-scale approaches using patch visualisation (attention mechanisms) and performing a medical study to reinforce the results of this paper would be of interest. Also, due to the low amount of data, high fluctuation in the classification results were observed. Increasing the data amount and repeating this experiment with variable amounts of training data to increase robustness of the findings would be a further, future goal.

%\section*{Acknowledgment}
%Blinded %That we do not forget
% This work was partially funded by the County of Salzburg under grant number FHS-2019-10-KIAMed.

\bibliography{citations}{}

\begin{thebibliography}{10}
\providecommand{\url}[1]{#1}
\csname url@samestyle\endcsname
\providecommand{\newblock}{\relax}
\providecommand{\bibinfo}[2]{#2}
\providecommand{\BIBentrySTDinterwordspacing}{\spaceskip=0pt\relax}
\providecommand{\BIBentryALTinterwordstretchfactor}{4}
\providecommand{\BIBentryALTinterwordspacing}{\spaceskip=\fontdimen2\font plus
\BIBentryALTinterwordstretchfactor\fontdimen3\font minus
  \fontdimen4\font\relax}
\providecommand{\BIBforeignlanguage}[2]{{%
\expandafter\ifx\csname l@#1\endcsname\relax
\typeout{** WARNING: IEEEtran.bst: No hyphenation pattern has been}%
\typeout{** loaded for the language `#1'. Using the pattern for}%
\typeout{** the default language instead.}%
\else
\language=\csname l@#1\endcsname
\fi
#2}}
\providecommand{\BIBdecl}{\relax}
\BIBdecl

\bibitem{siegel2020}
R.~L. Siegel, K.~D. Miller, and A.~Jemal, ``Cancer statistics, 2020,''
  \emph{CA: a cancer j. for clinicians}, vol.~70, no.~1, pp. 7--30, 2020.

\bibitem{tuttle2010thyroid}
R.~M. Tuttle, D.~W. Ball, D.~Byrd, R.~A. Dilawari, G.~M. Doherty, Q.-Y. Duh,
  H.~Ehya, W.~B. Farrar, R.~I. Haddad, F.~Kandeel \emph{et~al.}, ``Thyroid
  carcinoma,'' \emph{J. of the National Comprehensive Cancer Network}, vol.~8,
  no.~11, pp. 1228--1274, 2010.

\bibitem{gadermayr2021}
M.~Gadermayr, M.~Tschuchnig, L.~M. Stangassinger, C.~Kreutzer,
  S.~Couillard-Despres, G.~J. Oostingh, and A.~Hittmair, ``Frozen-to-paraffin:
  Categorization of histological frozen sections by the aid of paraffin
  sections and generative adversarial networks,'' in \emph{MICCAI: SASHIMI},
  2021, pp. 99--109.

\bibitem{hirokawa2002}
M.~Hirokawa, J.~A. Carney, J.~R. Goellner, R.~A. DeLellis, C.~S. Heffess,
  R.~Katoh, M.~Tsujimoto, and K.~Kakudo, ``Observer variation of encapsulated
  follicular lesions of the thyroid gland,'' \emph{The American j. of surgical
  pathology}, vol.~26, no.~11, pp. 1508--1514, 2002.

\bibitem{shah1992}
J.~P. Shah, T.~R. Loree, D.~Dharker, E.~W. Strong, C.~Begg, and V.~Vlamis,
  ``Prognostic factors in differentiated carcinoma of the thyroid gland,''
  \emph{The American j. of surgery}, vol. 164, no.~6, pp. 658--661, 1992.

\bibitem{duran2020}
L.~Duran-Lopez, J.~P. Dominguez-Morales, A.~F. Conde-Martin, S.~Vicente-Diaz,
  and A.~Linares-Barranco, ``Prometeo: A cnn-based computer-aided diagnosis
  system for wsi prostate cancer detection,'' \emph{IEEE Access}, vol.~8, pp.
  128\,613--128\,628, 2020.

\bibitem{li2018}
Z.~Li, Z.~Hu, J.~Xu, T.~Tan, H.~Chen, Z.~Duan, P.~Liu, J.~Tang, G.~Cai,
  Q.~Ouyang \emph{et~al.}, ``Computer-aided diagnosis of lung carcinoma using
  deep learning-a pilot study,'' \emph{arXiv preprint arXiv:1803.05471}, 2018.

\bibitem{bejnordi2017}
B.~E. Bejnordi, M.~Veta, P.~J. Van~Diest, B.~Van~Ginneken, N.~Karssemeijer,
  G.~Litjens, J.~A. Van Der~Laak, M.~Hermsen, Q.~F. Manson, M.~Balkenhol
  \emph{et~al.}, ``Diagnostic assessment of deep learning algorithms for
  detection of lymph node metastases in women with breast cancer,''
  \emph{Jama}, vol. 318, no.~22, pp. 2199--2210, 2017.

\bibitem{miyoshi2020}
H.~Miyoshi, K.~Sato, Y.~Kabeya, S.~Yonezawa, H.~Nakano, Y.~Takeuchi, I.~Ozawa,
  S.~Higo, E.~Yanagida, K.~Yamada \emph{et~al.}, ``Deep learning shows the
  capability of high-level computer-aided diagnosis in malignant lymphoma,''
  \emph{Lab. Investigation}, vol. 100, no.~10, pp. 1300--1310, 2020.

\bibitem{hou2016}
L.~Hou, D.~Samaras, T.~M. Kurc, Y.~Gao, J.~E. Davis, and J.~H. Saltz,
  ``Patch-based convolutional neural network for whole slide tissue image
  classification,'' in \emph{in CVPR}, 2016, pp. 2424--2433.

\bibitem{marini2021}
N.~Marini, S.~Ot{\'a}lora, F.~Ciompi, G.~Silvello, S.~Marchesin, S.~Vatrano,
  G.~Buttafuoco, M.~Atzori, and H.~M{\"u}ller, ``Multi-scale task multiple
  instance learning for the classification of digital pathology images with
  global annotations,'' in \emph{MICCAI: COMPAY}, 2021, pp. 170--181.

\bibitem{li2021}
B.~Li, Y.~Li, and K.~W. Eliceiri, ``Dual-stream multiple instance learning
  network for whole slide image classification with self-supervised contrastive
  learning,'' in \emph{CVPR}, 2021, pp. 14\,318--14\,328.

\bibitem{dimitriou2019}
N.~Dimitriou, O.~Arandjelovi{\'c}, and P.~D. Caie, ``Deep learning for whole
  slide image analysis: an overview,'' \emph{Frontiers in medicine}, vol.~6, p.
  264, 2019.

\end{thebibliography}
\bibliographystyle{IEEEtran}

\end{document}